# Microwave excitations and hysteretic magnetization dynamics of stripe domain films

Meihong Liu[1], Qiuyue Li[1], Chengkun Song[2], Hongmei Feng[2], Yawen Song[1], Lei Zhong[1], Lining Pan[2], Chenbo Zhao[2], Qiang Li[1], Jie Xu[1], Shandong Li[1], Jianbo Wang[2,3], Qingfang Liu[2], Derang Cao[1,2*]

[1]College of Physics, Center for Marine Observation and Communications, Qingdao University, 266071 Qingdao, China

[2]Key Laboratory for Magnetism and Magnetic Materials of the Ministry of Education, Lanzhou University, Lanzhou 730000, China

[3]Key Laboratory for Special Function Materials and Structural Design of the Ministry of the Education, Lanzhou University, Lanzhou 730000, China

## Abstract

FeNi films with the stripe domain (SD) pattern are prepared by electrodeposition and sputtering methods. The magnetic domain, static magnetic parameters, and quality factor, as well as dynamic properties of the two films, are respectively performed. The results show the magnetizations of the film were dependent on the direction of SD, and the rotation of the SD is lagging behind the magnetization reversal. The microwave properties of the SD emerge dynamic hysteresis before the saturation magnetic field. These microwave properties are selectively excited with acoustic mode, optical mode, and spin-wave mode. The frequency and intensity of different resonance modes of stripe domain are determined by the local magnetization. The magnetization variations and the rotation of SD of different modes are further illuminated by the micromagnetic simulation. The magnetic anisotropy and the resonance intensity of permeability of different modes were finally described by the modified resonance equations.

## 1. Introduction

During the past decades, extensive investigations are devoted to studying the static and dynamic properties of low-anisotropy ferromagnetic films both from the fundamental point of view and for numerous potential applications in magnetic storage media and microwave devices. These films can be distinguished by the type of anisotropy, i.e., in-plane and out-of-plane anisotropy [1]. Both the static magnetization configuration and the dynamic response are controlled by these key parameters. The most common studies of those films are characterized by an in-plane uniaxial anisotropy [2, 3], and another class of the films investigated corresponds to the ones exhibiting a weak perpendicular anisotropy [4]. It has been well ascertained that one of the remarkable characteristics is the existence of a stripe domain (SD) pattern for those films with such anisotropy [5, 6].

*Corresponding author: caodr@qdu.edu.cn

SD structure is firstly proposed in FeNi film in 1964 by Spain [7] and N. Saito *et al* [8, 9]. The researchers then provide the formation mechanism of SD which is related to the columnar grain structure induced perpendicular anisotropy of film [10-13]. It is now widely accepted [8, 11, 14, 15] that the physical origin of the SD structure is the competition between a moderate perpendicular magnetic anisotropy, exchange interaction, and the easy-plane dipole-dipole magnetostatic coupling energy. The domain structure is characterized by a periodic modulation of the up and down magnetization component as well as a Boloh- and Néel- type walls between two stripes. Such SD pattern is now evidenced in FeSi [16], Co [17], FeCo-based [18-20], NdCo [21], FeBSi [22], FePt [23], FeGa [6, 24], and FeNi [7, 25, 26] film. The film thickness of the SD is always above a critical value $t_c$ [14, 27]. The strength of SD is characterized by a quality factor, $Q=K_{\perp}/2\pi M_s^2$, which is defined as the ratio of the perpendicular anisotropy energy to magnetic dipole-dipole energy [1, 28]. $Q<1$, the magnetization tends to lie in the plane of the film; $Q>1$, the magnetization oscillates in the out of the plane.

Due to the up and down magnetic moment distributions of the SD film, the dynamic magnetization response and the anisotropy of the film become more fascinating and complex. The magnetic behavior of the films is in-plane isotropic but dependent on the magnetic history. This phenomenon further results in the multiple dynamic properties under the different magnetic configurations. Most researchers have studied the resonant model and the dynamic response of the SD structure film [1, 5, 20, 23, 28-44]. Particularly in detail, Acher *et al* [43] investigated the microwave permeability of SD film and mentioned the multiple resonance peaks to the spin-wave (SW). Vukadinovic *et al* [4, 14, 42] and Ebels *et al* [31] respectively proposed a ferromagnetic resonance (FMR) model and calculated the susceptibility spectra of SD structure films using a micromagnetic simulation. The results qualitatively explained that the source of different resonance peaks was related to the inside domain distribution of the SD. Chai *et al* [18, 45] and Tacchi *et al* [6, 24] focused their investigations on the rotational anisotropy of SD film by measuring the magnetic spectra dependent on the in-plane angle between SD and magnetic field. Recently, Tacchi *et al* [46] also phenomenologically compared the different magnetization dynamic modes of SD film via the Brillouin light scattering (BLS) and FMR techniques. The above-previous studies conclude the very interesting results that the multiple resonances of magnetic excitation in the SD film depend on the specific pumping field configurations concerning the stripe direction and magnetic field. The frequencies dependence of these modes reflects the changes of the local and internal anisotropy fields of SD structure. In this case, however, there are still some confusions that need to be solved. In another word, the process and relation between the SD rotation and its inside magnetization reversal as well as the quantitative anisotropy are not yet clear. It is necessary to further propose the evolution processes of SD films under different resonant modes.

In this work, the typical parameters of the FeNi SD film were first studied and given. By comparing the static magnetization of SD film, the dynamic response was measured under different pumping- or applied magnetic- field configurations concerning both the stripe and magnetization of the SD film. The results discuss the variation of the domain, magnetization distribution, hysteresis, resonant frequency, permeability, and anisotropy of the SD film in those configurations. The process of the stripe rotation and its inside magnetization reversal is further demonstrated by the micromagnetic simulation. This study showed a clear evolution of the magnetization, stripe, and the relevant anisotropy of SD films. The results provided a favorable point to further understand the magnetization dynamics of the SD.

## 2. Experiment process and method

The films were deposited on Si (100) substrates and indium tin oxide (ITO) substrates by RF magnetron sputtering and electrodeposition method respectively. The component is $Fe_{65}Ni_{35}$ for electrodeposited film and $Fe_{20}Ni_{80}$ for sputtered film and the corresponding thicknesses $t_{film}$ of the two films are 500±15 nm and 200±8 nm respectively. We note that the composition of electrodeposited film and the sputtered film is different. Our previous composition-dependent results [47] indicate FeNi films exhibit good acoustic and optic mode resonances properties when Ni at.% is between 25% and 38% for the electrodeposited film. This does not affect the research content involved in this work. The deposition rate of electrodeposited film and the sputtered film is about 50 nm/min and 3 nm/min respectively. For the magnetron sputtering technique, the target was Permalloy, the base pressure of the vacuum chamber was better than $5\times10^{-5}$ Pa, and the sputtering power and Ar pressure were 120 W and 0.4 Pa. For the electrodeposition method, the deposition potential was -1.2 V, and the contents of electrolyte were composed of $FeSO_4\cdot7H_2O$ (0.05 mol/L), $NiSO_4\cdot7H_2O$ (0.05 mol/L), $H_3BO_3$ (0.5 mol/L), $C_6H_8O_6$ (1 g/L), $C_2H_5NO_2$ (2 g/L) and $C_7H_5O_3NS$ (2 g/L), which was also afforded in our previous work [48].

Magnetic force microscopy (MFM, Asylum Research MFP3D) was used to study the domain structures of the films. The compositions of the samples were identified by an energy-dispersive X-ray spectrometer (EDX, Hitachi S-4800). The in-plane hysteresis loops and magnetization curves of samples were measured by vibrating sample magnetometer (VSM, Lakeshore 7304) and B-H Loop analyzer (Riken Denshi, BHV-30S) at room temperature respectively. The ferromagnetic resonances (FMR) of the films were performed by electron spin resonance (ESR, JEOL, JESFA300) with an X-band spectrometer at 9 GHz. The permeability spectra of all films were measured by a vector network analyzer (VNA, Agilent E8363B) method from 100 MHz to 9 GHz [49].

## 3. Results and discussion

### A. Basic structure and static magnetic properties of SD film

Figure 1 (a-b) shows the zero-field MFM images of electrodeposited and sputtered FeNi Dark and bright regions (domains) represent an out-of-plane component of the magnetization with upward or downward directions, respectively, and between the two close stripes is the domain wall. The average SD width of the two films is 167±13 nm (electrodeposition) and 143±5 nm (sputtering) respectively. It can be observed from the pictures that the electrodeposited film presents a dispersive SD structure, while the SD pattern is more uniform and clear in the sputtered film. The SD pattern of the film shows good regularity and homogeneity domain i.e., evenly spaced and parallel stripe distributions with a single and strong Fast Fourier transform peak. This quantitative definition has been given in our previous work [50]. Our previous work of SD patterns prepared by these two ways always shows such different domain results [48, 51], and the results are related to the film quality of the two methods. The film prepared by sputtered usually has better quality than electrodeposited film. It has been proved that the obvious and uniform SD pattern reveals a strong and clear resonant mode [50]. As mentioned in the introduction, the level of SD is characterized by a quality factor $Q$ ($Q=K_\perp/2\pi M_s^2$). The perpendicular anisotropy constant $K_\perp$ and saturation magnetization $4\pi M_s$ can be obtained by measuring the FMR. The inset of figure 1 (e-f) is the out-of-plane angular of the magnetic field ($\theta_H$) dependence of the out-of-plane FMR field ($H_R$). By fitting the relation between $\theta_H$ and $H_R$ [48, 52], the $4\pi M_s$ and $K_\perp$ are 13.1 kGs and $6.3\times10^5$ erg/cm$^3$ for electrodeposited film and 0.9 kGs and $5.5\times10^5$ erg/cm$^3$ for sputtered film respectively. The calculated $Q$ is 0.09 and 0.17 for electrodeposited and sputtered films respectively. The sputtered film shows a larger $Q$ than that of electrodeposited film. The results confirm the obvious SD pattern of sputtered film than electrodeposited film. We note that when other parameters are fixed, the SD width decreases with the increasing saturation magnetization, while it increases with the increasing thickness of the film. The same conclusion can be found in previous publications [1, 12, 53, 54]. Such results can be also easily seen from Equation (S1) of domain width in our Supplemental Material. For the quality factor, it is complicated and is related to the saturation magnetization, thickness (demagnetization), and perpendicular anisotropy field. This is the reason that we observed the larger perpendicular anisotropy but smaller quality factor in the electrodeposited film. In addition, according to the above values of $4\pi M_s$ and $K_\perp$, the domain width was also calculated by the theory expression [55, 56]. See the Supplemental Material for details process. The calculated SD width of two films using Eq. (S1) is about 154 nm for electrodeposited film and 144 nm for the sputtered film, which is comparable with the MFM value. It can be seen the error between the measured value and the calculated value for the electrodeposited film is larger. This may be due to the dispersive and inhomogeneous SD pattern in electrodeposited film, which causes a large measurement error.

Figure 1 (c-d) displays the in-plane

hysteresis loops recorded on the electrodeposited and sputtered FeNi SD films. Both the two loops are characterized by a linear decrease of magnetization from its saturation field to a moderate remanence. This shape is the typical loop for thin films exhibiting a SD structure [47]. The hysteresis loops of the SD films are in-plane isotropic wherever the orientation of the in-plane applied magnetic field is (See figure S3 of Supplemental Material). This reflects the so-called rotatable anisotropy effect in SD [1]. Figure 1 (e-f) are the amplifying results of the first and fourth quadrant hysteresis loops of (c) and (d) respectively. This step was used to describe the following VNA measurement conveniently. The letters $a_1$~$j_1$ of the upper solid line represent the applied magnetic field $H$ changing from 180 Oe to 0 Oe, while the letters $j_2$~$a_2$ mean that the $H$ is from 0 Oe to 180 Oe. The step size of $H$ is 20 Oe. It can be observed that the upper and lower lines are dividing, and are starting to approach when $H$ is larger than 70 Oe for electrodeposited film and 90 Oe for the sputtered film. We note that, during the measurement, both the intensity and direction of $H$ for points $a_1$ and $a_2$, $b_1$ and $b_2$, ⋯ $j_1$ and $j_2$, are respectively the same, but the relevant magnetizations (both the intensity and direction) are different. We deduce that such result is related to the stripe or internal magnetizations of SD. Our following study will systematically discuss this process.

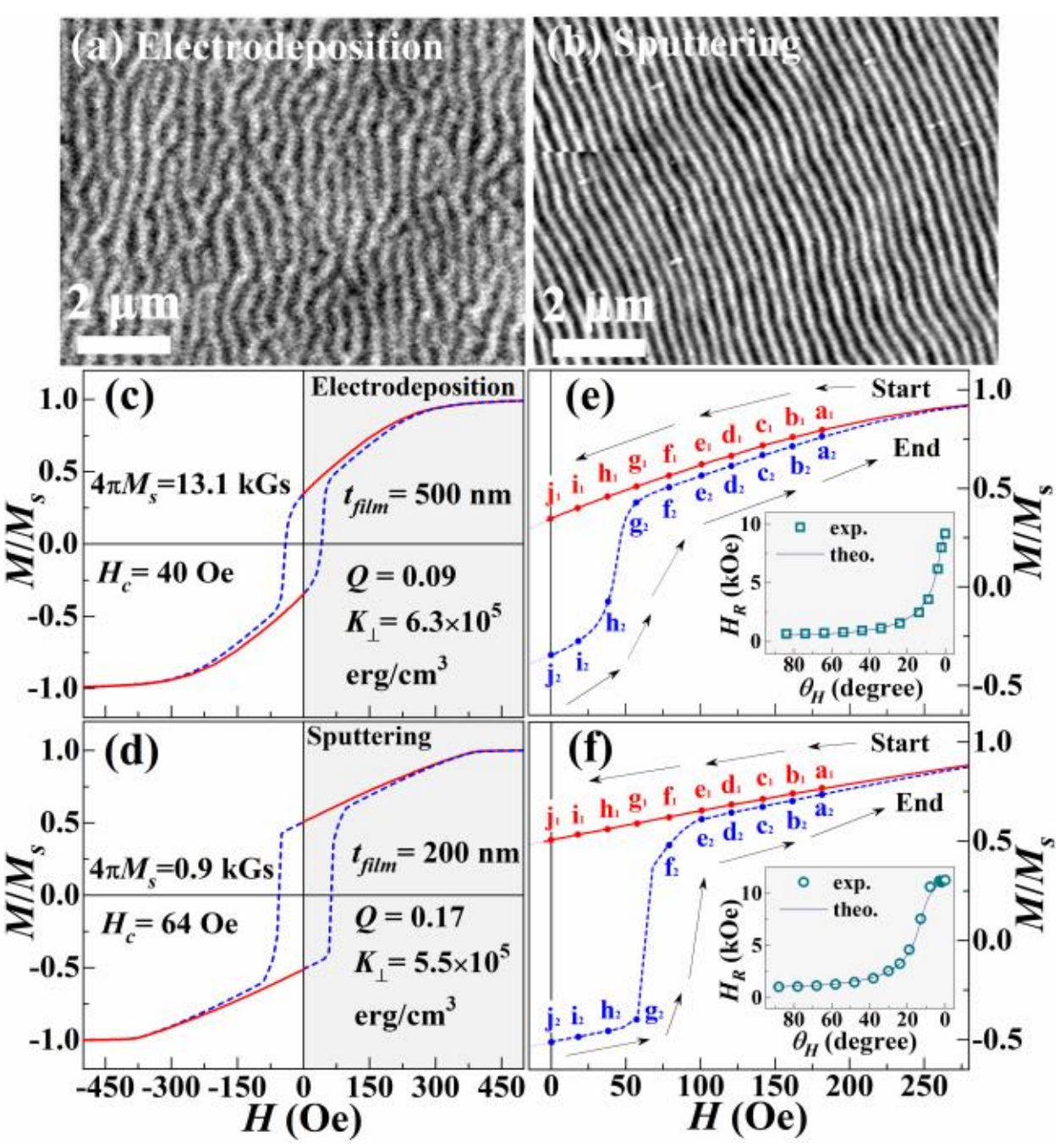

Figure 1. (a-b) Zero-field MFM images of electrodeposited and sputtered FeNi films. (c-d) In-plane hysteresis loops of the two sample films. (e-f) the amplifying loops of the first and fourth quadrant. The letters $a_1$~$j_1$ of the upper solid line represent the applied magnetic field $H$ changing from 180 Oe to 0 Oe, while the letters $j_2$~$a_2$ mean the $H$ is from 0 Oe to 180 Oe. The step size of $H$ is 20 Oe. The inset of the figure (e-f) is the corresponding out-of-plane angular of the magnetic field ($\theta_H$) dependence of the out-of-plane FMR field ($H_R$).

For the SD film, as well known, it is possible to select the easy direction of the magnetization by the application of a sufficiently large in-plane magnetic field along this direction and then remove it. To further study the static rotation of the stripe and magnetization, we further studied the in-plane magnetization curves of SD films at different orientations. The results are shown in figure 2. For each measurement, the direction of the SD is fixed parallel to the $x$-axis, and $H$ is applied along the $\pm x$-axis or $y$-axis. Thus, three different situations are carried out:

i) $H$ is applied along the $x$-axis, i.e., parallel to the direction of SD and $M$ (figure 2c).

ii) $H$ is applied along $-x$-axis, i.e., parallel to the direction of SD and anti-parallel to the direction of $M$ (figure 2d).

iii) $H$ is applied along the $y$-axis, i.e., perpendicular to the direction of SD and $M$. $M$ has two orientations here, but the results of the two situations are the same (figure 2e).

The in-plane magnetization curves of electrodeposited and sputtered SD films are shown in figure 2 (a-b) respectively. The following rules are declared from the curves:

i) Three different curves are observed and highly related to the initial direction of $H$ and SD.

ii) For $H$//SD, the initial $4\pi M$ value (when $H$ is zero) of curve measured at situation $H\uparrow//M\uparrow$ and $H\downarrow//M\uparrow$ is the same, but the direction is opposite.

iii) For $H\perp$SD, the curves measured at situation $H\uparrow\perp M\uparrow$ and $H\downarrow\perp M\uparrow$ are the same. Here we only present the results of $H\uparrow\perp M\uparrow$. Their initial $4\pi M$ (when $H$ is zero) is closed to zero. This is due to the near-zero component of magnetization in the $y$-axis at this time.

iv) For $H$//SD, the $4\pi M$ of the $H\uparrow//M\uparrow$ and $H\downarrow//M\uparrow$ show a linear increase with the improvement of $H$ (Note: for $H\downarrow//M\uparrow$, this linear increase happens as $H$ exceeds a reversal field $H_{rev}$). The two curves are similar to the VSM loops of figure 1 (e-f). The $H_c$ obtained from the curves is consistent with the VSM loops.

v) For $H\perp$SD, the $4\pi M$ presents a linear increase first when $H$ is less than a transition field $H_{tra}$. $H_{tra}$ is about 72 Oe for electrodeposited film and 123 Oe for the sputtered film. The $4\pi M$ then continues to increase but with a reduced slope as $H$ further increases.

vi) The $4\pi M$ of the three situations starts to overlap when the $H$ is larger than $H_{tra}$, and finally reaches saturation.

The results indicate that the magnetization process of SD film is strongly dependent on the direction of SD and its inside magnetization configuration.

When $H$//SD, for the situation $H\uparrow//M\uparrow$, the $M$ inside SD is always parallel to the direction of $H$. For the situation $H\downarrow//M\uparrow$ the $M$ is anti-parallel at the beginning of the curve, and starts rotating to the direction of $H$ when $H$ exceeds $H_c$. They finally point to the direction of $H$ when $H$ is larger than $H_{rev}$. We note that the direction of SD is still parallel to $H$ during the two processes ($H\uparrow//M\uparrow$ and $H\downarrow//M\uparrow$).

When $H\perp$SD, the $M$ inside SD is also perpendicular to $H$. Both the $M$ and SD start to rotate with the increase of $H$, but their relevant critical-rotational field is different. The $M$ begins to move once the $H$ is applied, while the SD rotates when $H$ is near $H_{tra}$. These results will be demonstrated in the following micromagnetic simulations.

The above results reveal that the rotation of SD is hysteretic compared with its inside magnetization. This is the origin of a rotational anisotropy in SD film. Some researchers call this anisotropy a pseudo-anisotropy [16]. This is due to the anisotropy depends on the direction of SD and is controlled by the external magnetic field.

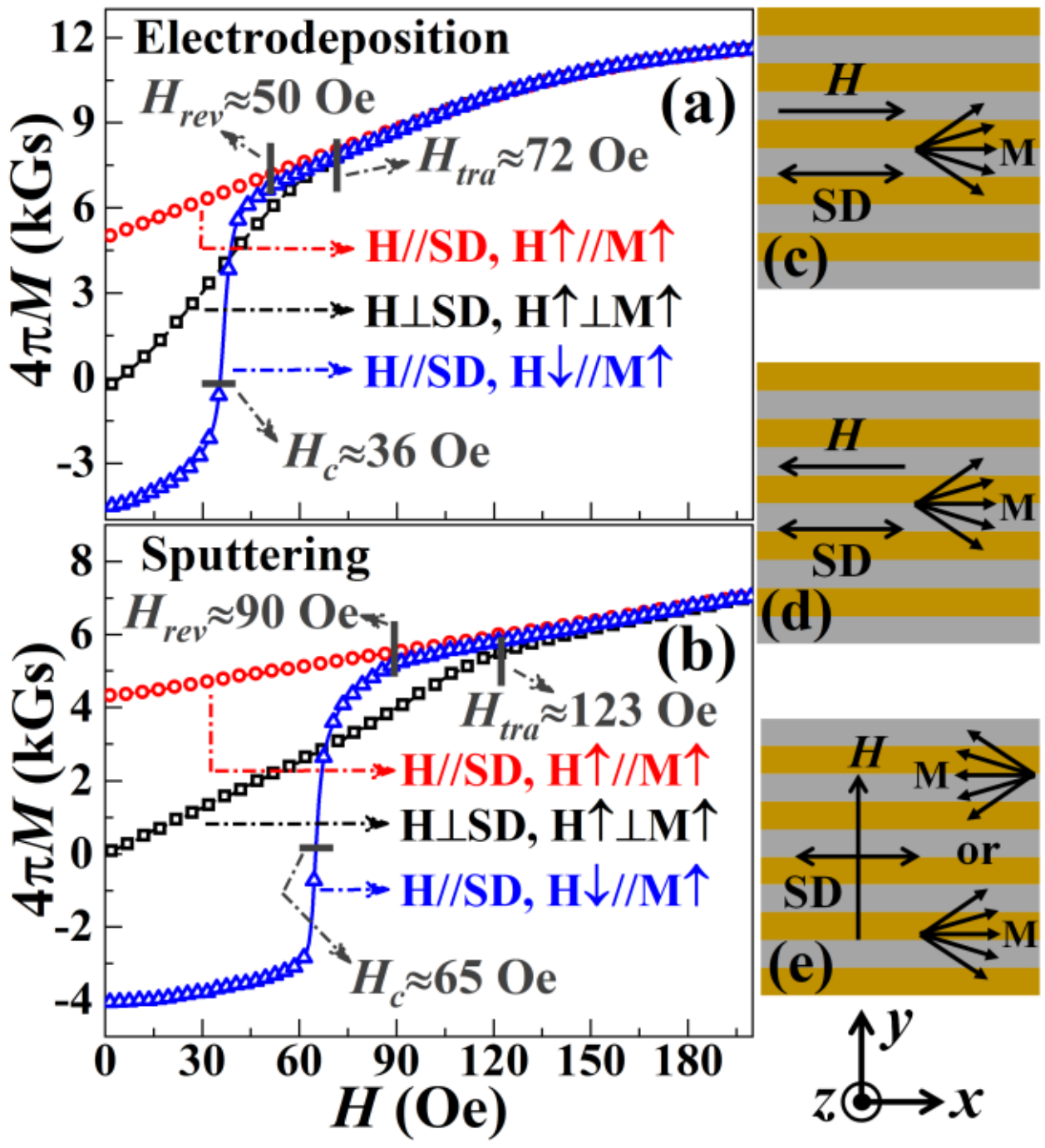

Figure 2. (a-b) In-plane magnetization curves of electrodeposited and sputtered FeNi SD films. The curves were measured at the different directions of SD, magnetization *M*, and applied magnetic field *H*. The red circle line is the situation *H*//SD and *H*↑//*M*↑; the blue triangle line is the situation *H*//SD and *H*↓//*M*↑; the black square line is the situation *H*⊥SD and *H*↑⊥*M*↑. (c-e) The orientation schematic diagram of the SD, *M*, and *H* during measurement. The solid line arrows are the direction of SD, *M*, and *H* respectively.

## B. Dynamic magnetic properties of the film

The dynamic magnetic spectra of films were performed by VNA with an applied magnetic field *H*. The direction of the microwave magnetic field $h_{rf}$ is fixed orthogonality to *H*. The results are shown in figure 3. Figure 3 (a) and (c) shows the imaginary permeability spectra of electrodeposited and sputtered FeNi films at different *H*. The application process of the *H* is the same as VSM measurement. The letters, $a_1$~$j_1$ and $a_2$~$j_2$, are corresponding to the application of *H* in VSM. Figure 3 (b-d) shows the color-coded imaginary permeability spectra of the two FeNi films as a function of the applied field. One can observe the following comments from figure 3:

i) In figure 3, the application of *H* is the same as the first and fourth quadrant of VSM measurement. Similarly, we can get the same spectra when the application procedure of *H* accords with the second and third quadrant of VSM measurement.

ii) Both the value and direction of *H* for $a_1$~$j_1$ (180~0 Oe) and $j_2$~$a_2$ (0~180 Oe) are the same during the permeability spectra measurement. However, the corresponding spectra are different and replaced by a hysteretic behavior.

iii) The color-coded permeability spectra are asymmetric about zero fields, and an inflection appears in the range of 15~45 Oe for electrodeposited film and 45~80 Oe for the sputtered film. It is approximately symmetrical when *H* exceeds about 140 Oe for electrodeposited films and 120 Oe for the sputtered film.

iv) In the asymmetric region, except zero fields both the frequency and intensity of spectra are different when *H* is in a small range (corresponding to field at letters e, f, g, h, and i). It can be seen that the corresponding spectra in figure 3 (a) and (c) are non-overlapping.

v) The law of two color-coded magnetic spectra (figure 3b and 3d) is similar to the VSM loops (figure 2c-d). The value of the magnetic field in the inflection region of the spectra is close to their $H_c$ ($H_c$ obtained from VSM are 40 Oe for electrodeposited film and 64 Oe for sputtered film respectively). The results indicate the asymmetric spectra are related to their current magnetization state.

vi) The spectra of the sputtered film with higher frequency (figure 3d) are attributed to spin-

wave (SW) [4, 40, 43], but it is not observed in the electrodeposited film. This may be due to the weak SW being overlapped with the broad resonance linewidth of electrodeposited film or the SW cannot be excited in the low-regular SD pattern [50] of the electrodeposited film (see MFM results).

The above results indicate that this asymmetric spectrum (both the frequency and intensity) is related to the magnetization and reversal magnetization along with SD film. The $H$-dependent frequency and intensity of spectra will demonstrate in Sec. 3D. As a comparison, the same measurement is performed in $Fe_{45}Co_{55}$ and $Fe_{20}Ni_{80}$ films with the in-plane uniaxial anisotropy, and the results are shown in figure S1 of Supplemental Material. The results indicate that this dynamic hysteretic behavior is almost disappeared. Their corresponding spectra are overlapping well, and show an ideal symmetric color-coded spectrum about zero fields. Acher *et al* [3] previously demonstrated that the soft magnetic thin film with in-plane anisotropy showed the dynamic hysteretic behavior when the applied field was smaller than its $H_c$. The applied field was very small (<3 Oe) when dynamic hysteretic behavior happened. This effect can be negligible in our in-plane films, where the applied field step (10 Oe) is relatively large. However, the dynamic hysteretic behavior in SD film is more obvious. This dynamic hysteretic behavior was also observed in the strong exchange biased film [57, 58]. The film always showed the asymmetric loops about zero fields.

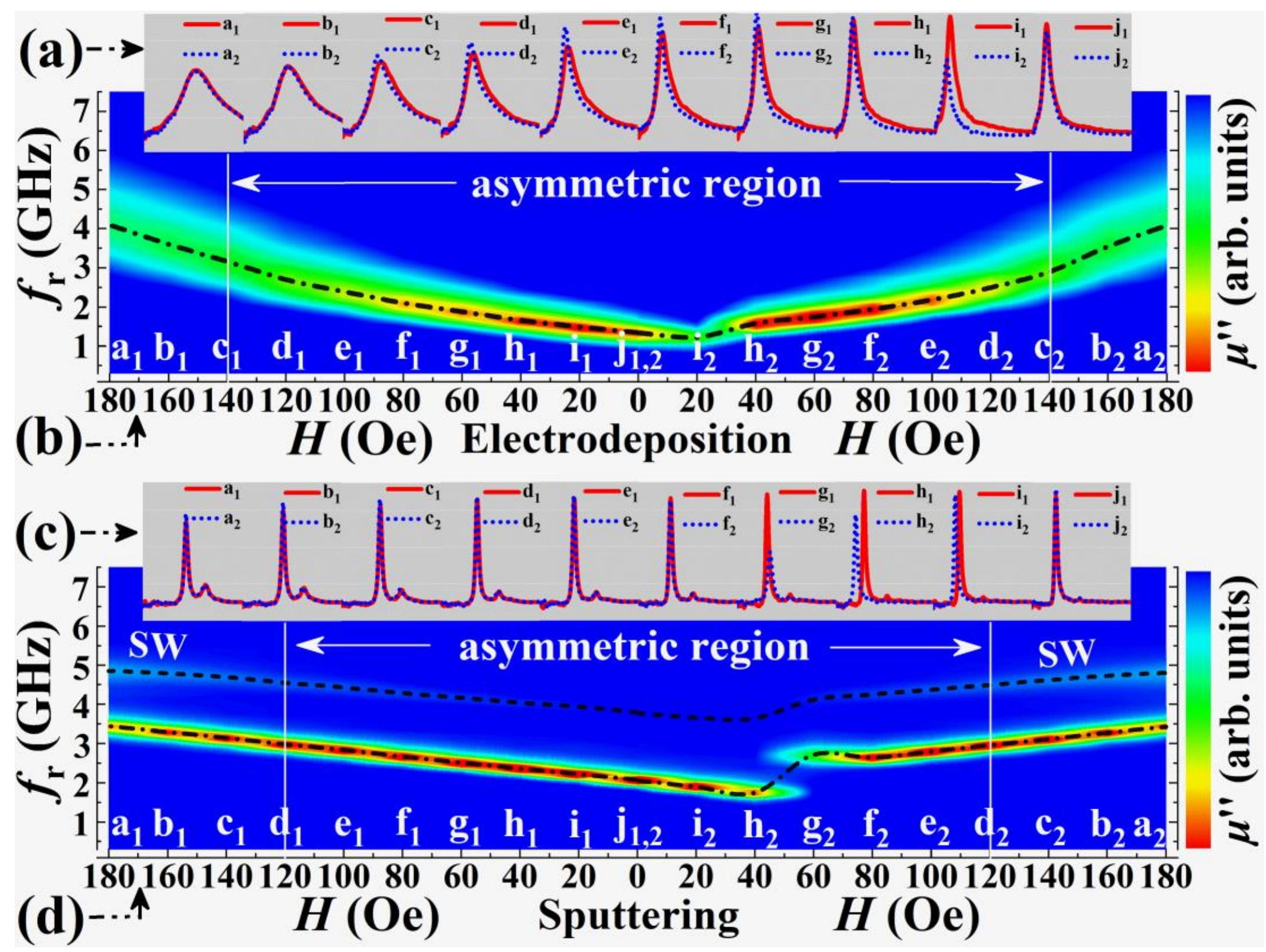

Figure 3. (a) and (c) Imaginary permeability spectra of electrodeposited and sputtered FeNi films measured at different $H$. (b) and (d) Color-coded imaginary permeability spectra of the two FeNi films as a function of $H$; the short dash-dot is the maximum frequency value of permeability spectra. The letters ($a_1$~$j_1$ and $a_2$~$j_2$) represent the application of $H$ (both intensity and direction), and it is the same as VSM measurement (figure 2c-d).

To clarify the dynamic hysteretic behavior and the dynamic properties corresponding to the magnetization curves, the magnetic spectra dependence of the applied magnetic field $H$ and microwave magnetic field $h_{rf}$ are further demonstrated in figure 4. The $h_{rf}$ is fixed orthogonality to $H$ during the measurement. The application of $H$ is the same as the magnetization curves of figure 2 (a-b). The directions of the $M$ and SD are orientated by the application of a sufficiently large in-plane $H$ and then remove it. The following comments can be obtained:

i) Two modes are observed in figure 4. When the direction of $h_{rf}$ is perpendicular to SD ($h_{rf}\perp$SD), the measurement is the acoustic mode (AM) with in-phase precession. When the direction of $h_{rf}$ is parallel to SD ($h_{rf}$//SD), the measurement is the optical mode (OM) with out-of-phase precession [23, 31]. The resonance frequency and intensity of AM and OM are different. AM has a lower resonance frequency but higher intensity while OM has a higher resonance frequency but lower intensity, In AM and OM, they have their own corresponding SW resonance.

ii) For all AMs (figure 4a, 4b, 4d, and 4e), the resonant peak at the low frequency is in accordance with the classical in-plane uniform precession mode [43]. The spectra with a higher frequency in the sputtered film is the SW. The SW of AM originates from the spin resonance inside the Bloch-type domain walls and the spin resonance at the film surface [4, 14, 31].

iii) For the AM, the frequency $f_r$ of $H\uparrow//M\uparrow$ (figure 4a and 4d) increases with the improvement of $H$, while it decreases first and the increase for $H\downarrow//M\uparrow$(figure 4 b and 4e). This is related to the opposite initial direction between the $H$ and $M$ when $H$ is less than their $H_c$.

iv) For the OM (figure 4c and 4f), the intensity of OM weakens when compared with AM. The OM floats at about 3 GHz for electrodeposited film and about 5.5 GHz for the sputtered film. The AM emerges when $H$ is larger than about 40 Oe for electrodeposited film and 60 Oe for sputtered film, and their $f_r$ is increasing with the increased $H$. The higher $f_r$ (>6 GHz) is SW. The SW of OM originates from the spin resonance inside the Néel-type like domain and the spin resonance in the upper and lower part of volume [4, 14, 31]. The transitions of the OM and AM are related to the rotation of the SD with the increase of the $H$. This changes the relative direction between the $h_{rf}$ and SD, and thus, different resonant modes are excited at the same time. It can be seen multiple resonance peaks (such mix resonance peaks including SW, OM, AM, domain wall resonance) in the range of $H$ = 90 Oe~150 Oe, and it is difficult to clearly distinguish these peaks. These resonances contain the contributions from magnetizations inside the domain, domain wall, and the flux closure cap [4, 14, 31, 46]. The later micromagnetic simulations also reveal a complicated domain structure. All the modes will become the in-plane uniform precession mode when SD disappears.

v) For the AMs of $H\uparrow//M\uparrow$ and $H\downarrow//M\uparrow$ in figure 4 (a-b) and (d-e), the results are similar with the law of figure 3 (b) and (d) respectively.

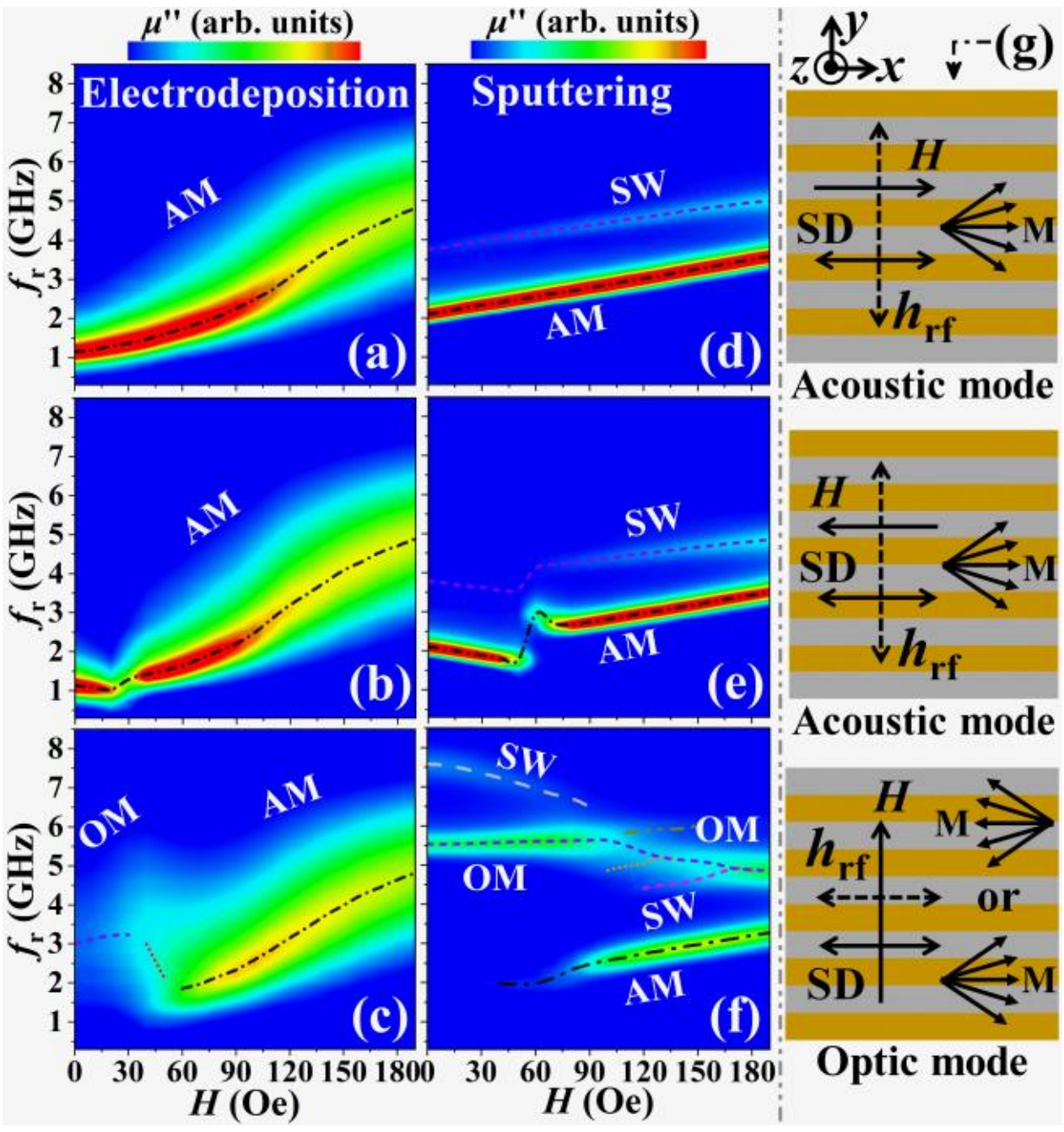

Figure 4. (a-f) Color-coded imaginary permeability spectra corresponding to the different directions of the electrodeposited film (a-c) and sputtered film (d-f) as a function of $H$; the short dash-dot is the frequency value of permeability spectra. The application of the magnetic field of the magnetic spectra measurement is the same as the magnetization curves of figure 2 (a-b). (g) The orientation schematic diagram of the SD, $M$, $H$, and $h_{rf}$ during measurement. The solid line arrows are the direction of SD, $M$, and $H$ respectively, and the dash line arrow is the direction of $h_{rf}$. The $h_{rf}$ is fixed orthogonality to $H$ during the measurement.

## C. Micromagnetic simulations

Micromagnetic simulations have been performed using the graphics processing unit (GPU) accelerated software Mumax3 [59]. In micromagnetic simulations, we assumed effective homogeneous material parameters in the whole structure. The simulations were relaxed to compute dispersion relations for the uniformly magnetized structure and the stable SD pattern. The parameters in the simulations are as follows [40]. The saturation magnetization was set to $4\pi M_s$ = 10 kGs. The exchange stiffness constant and the out-of-plane perpendicular anisotropy were $A_{ex}$=7.2×10$^{-7}$ ergs/cm and $K_\perp$ = 5.0×10$^5$ ergs/cm$^3$, respectively. The values of the quality factor were then calculated with $Q$= $K_\perp/2\pi M_s$=0.13. The total simulated size is 1000×1000×200 nm$^3$ and was discretized into cells with the dimensions of 5×5×5 nm$^3$. A relaxation method is used iteratively in order to achieve the random alignment of the magnetization vector in the minimum energy.

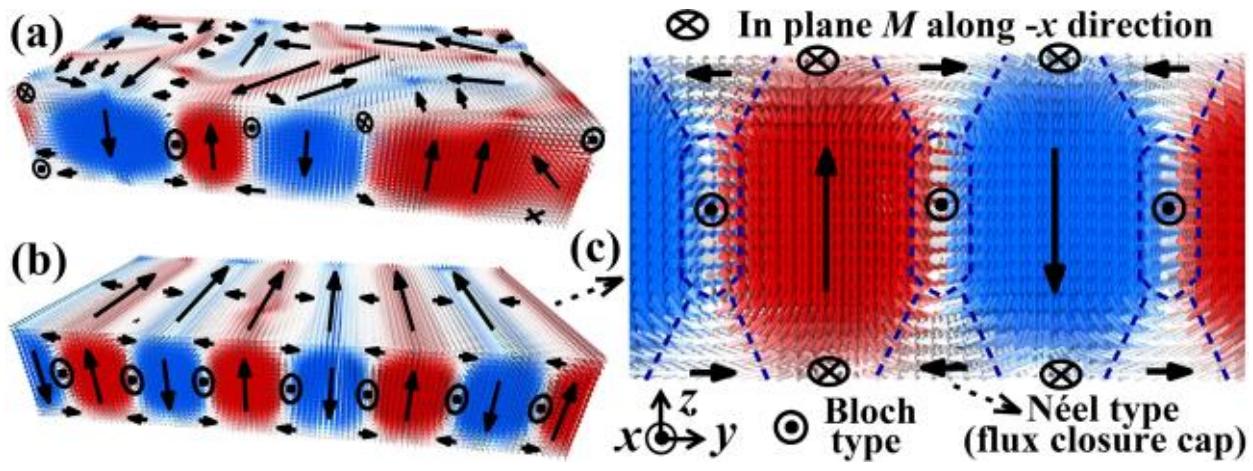

Figure 5. 3D micromagnetic simulation result for square structures 1000×1000×200 nm$^3$. The color scale is adopted for the magnetization component $M_z$ along the $z$-axis. Red regions have positive $M_z$, teal regions have negative $M_z$, and white regions have $M_z$ = 0. (a) Initial state; (b) Remanence state. (c) Cross-section of magnetization at the remanence state in one period. The front plane of (a-b) shows the equilibrium magnetization of the film plane ($yz$ plane) at half of the film. The black arrows indicate the direction of magnetization.

Figure 5 presents the magnetization distribution simulated at the initial state and remanence state. The arrows represent the projection of the magnetization in the plane ($x$, $y$) while the component $M_z$ is given by a teal-white-red color-code map. Figure 5 (a) shows a random stripe distribution at the initial state while figure 5 (b-c) presents the periodic arranging stripes at the remanence state [figure 5 (c) is the amplifying image]. These distribution diagrams can be more distinctly seen in figure S2 of Supplemental Material by 2D plotted pictures. At remanence, the following conclusion is deserved:

i) It can be observed from figure 5 (b) that the basic domains in SD show an alternately up and down (along the *z*-axis) magnetization concerning the film plane. The result is in reasonable agreement with the experimental MFM images. At the film surface, these up and down magnetizations are oriented along the *x*-axis (the direction of the magnetizing history) with slightly tilting concerning the film surface due to the large demagnetization energy.

ii) The middle of the adjacent stripe is separated by Bloch-type domain walls magnetized along the +*x*-axis. These walls lead to SD paralleling the direction of the saturation field and presenting the periodicity.

iii) Due to the moderate value of the out-of-plane perpendicular anisotropy $K_{\perp}$, a flux closure cap of Néel-type like domains consists of a region near the film surface and in-plane magnetized along the *y*-axis. The flux closure cap domain is the characteristic of the weak SD regime, and the size of the region is in accordance with the values of the quality factor $Q$, domain width $w$, and critical thickness $t_c$ [46]. Since SD has an equal size and is alternately magnetized, it can be seen (figure 5b) that the sum of $M_y$ at remanence is zero. This is in agreement with the magnetization curves of figure 2 (situation $H\perp$SD).

To validate the experimental results and to understand the physical characteristics of the rotation of SD and magnetization, the micromagnetic simulations are further carried out with a different direction of SD, $M$, and $H$ respectively. The film is fixed at remanence before simulation. The $H$ is set at the same direction with the measurement of magnetization curves (figure 2), i.e., three situations: $H$//SD and $H\uparrow$//$M\uparrow$, $H$//SD and $H\downarrow$//$M\uparrow$, $H\perp$SD and $H\uparrow\perp M\uparrow$ or $H\downarrow\perp M\uparrow$. We note that for $H\perp$SD, the results of $H\uparrow\perp M\uparrow$ or $H\downarrow\perp M\uparrow$ are the same, here we only give the results of $H\uparrow\perp M\uparrow$. The simulation results are shown in figure 6 (top view) and figure 7 (cross-section). The intensity of $H$ has been increased from 0 to 3000 Oe by steps of 100 Oe.

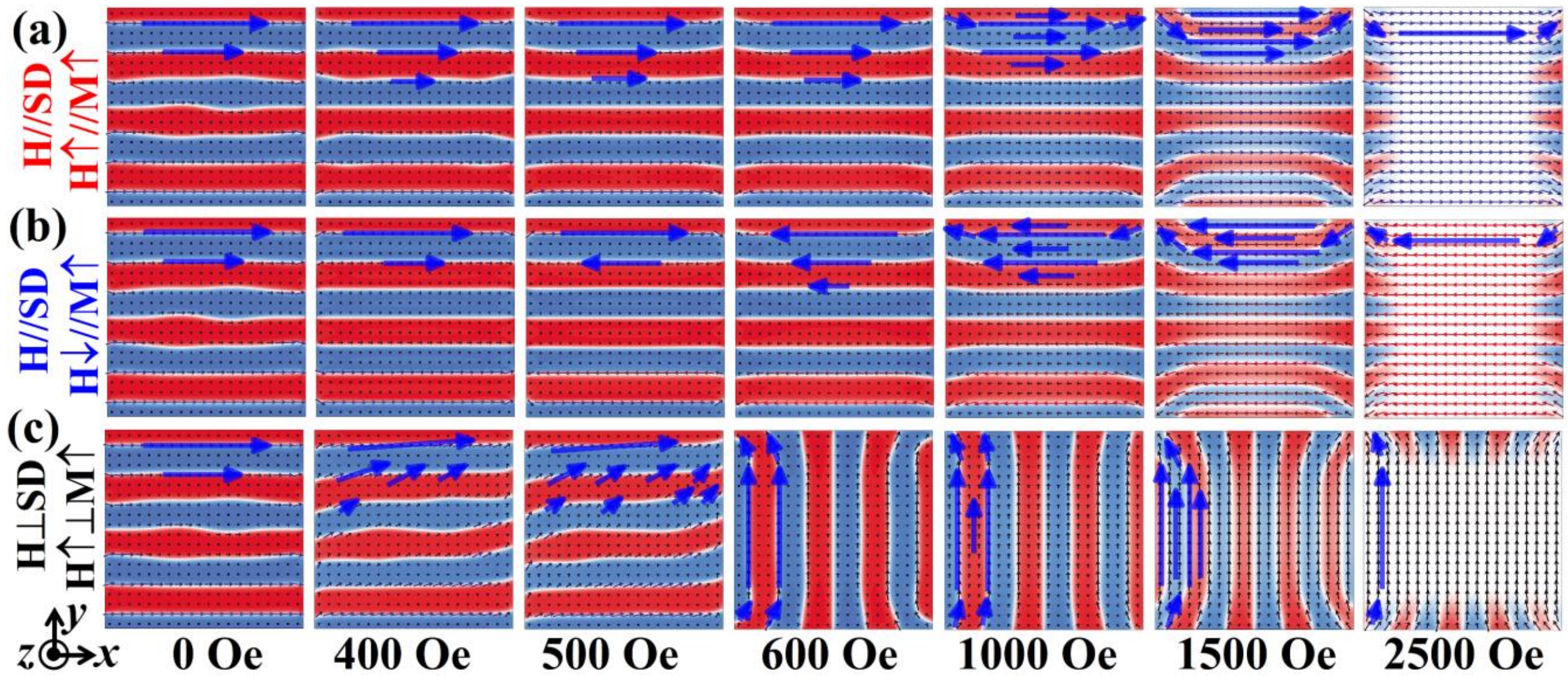

Figure 6. Top view of the film plane (*xy* plane) for the micromagnetic simulations at the various directions of SD, *M*, and *H*. The images show the equilibrium magnetization of the film plane at half of the thickness. Before applying *H*, the initial state of films is remanence and the direction of SD and *M* is parallel to the *x*-axis. The simulation and color scale are the same as in figure 5. The large and blue arrows guide the direction of the in-plane (*xy* plane) magnetization in one stripe period. (a) *H*//SD and *H*↑//*M*↑, *H* is applied along the +*x*-axis. (b) *H*//SD and *H*↓//*M*↑, *H* is applied along −*x*-axis. (c) *H*⊥SD and *H*↑⊥*M*↑, *H* is applied along +*y*-axis. The blue arrow is to clarify the change in the local magnetization directions during the three different magnetization processes. The length of the blue represents the quantity of the local magnetization here along this direction.

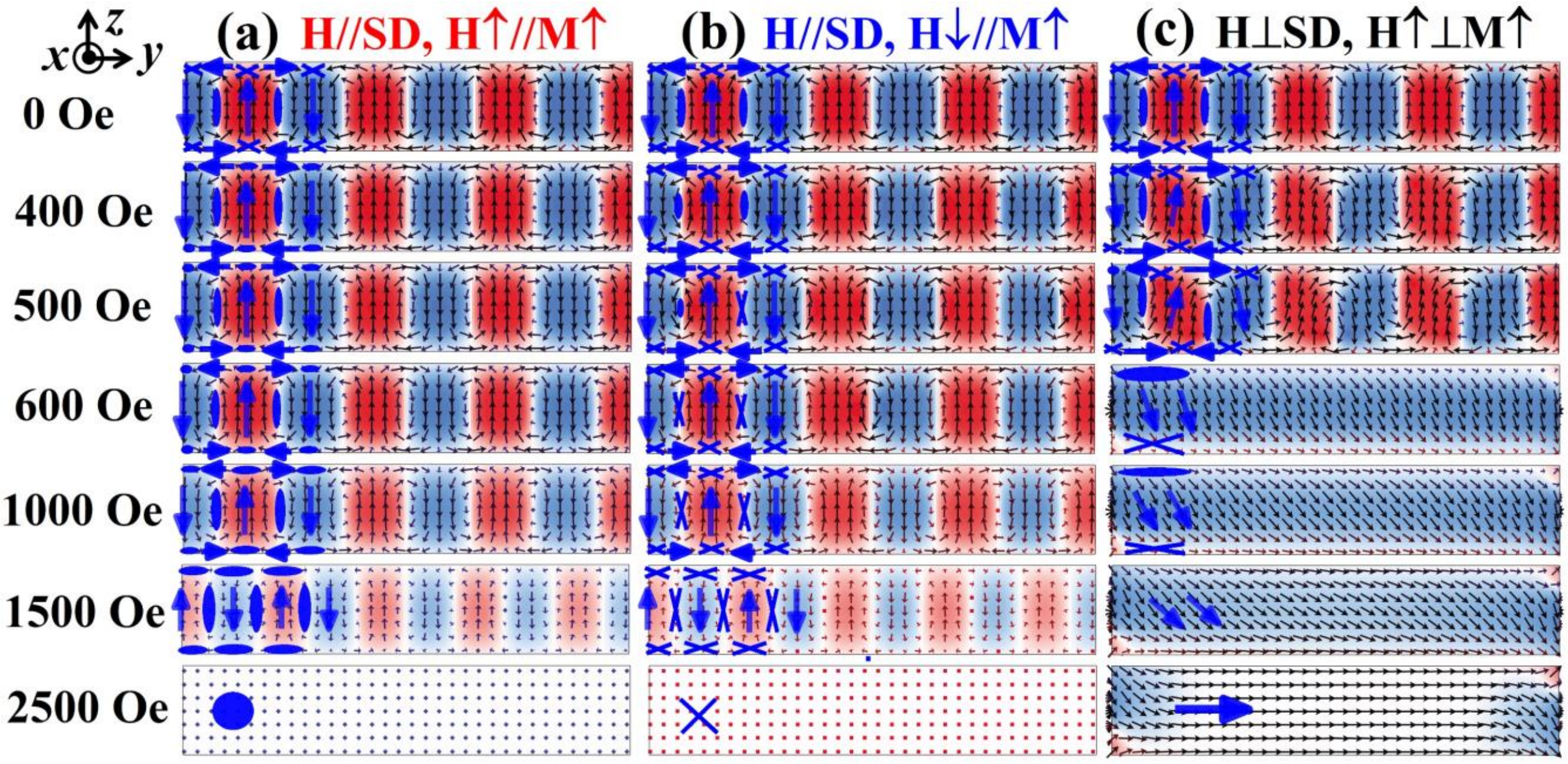

Figure 7. Cross-section of magnetization (*yz* plane) for the micromagnetic simulations at the various directions of SD, *M*, and *H*. The images show the equilibrium magnetization of the film plane at half of the film plane. The simulation and color scale are the same as figure 5 and figure 6. The large and blue arrows guide the direction of the cross-section (*yz* plane) magnetization in one stripe period; "●" represents the magnetization is "out" in the region,

while “×” represents the magnetization is “in” in the region. (a) $H$//SD and $H$↑//$M$↑, $H$ is applied along +$x$-axis. (b) $H$//SD and $H$↓//$M$↑, $H$ is applied along the –$x$-axis. (c) $H$⊥SD and $H$↑⊥$M$↑, $H$ is applied along +$y$-axis. The blue arrow is to clarify the change in the local magnetization directions during the three different magnetization processes. The length of the blue represents the quantity of the local magnetization here along this direction.

The detailed description of figure 6 and figure 7 are displayed as flown:

I) When $H$ is zero, the pictures are the same as figure 5 (b-c) and figure S2 (c-d). The component $M_x$ comes from the magnetization along the +$x$-axis in Bloch-type domain wall and a tilted in-plane magnetization along –$x$-axis near the film surface. $M_y$ is composed of the magnetization along the $y$-axis at a flux closure cap of Néel-type like domains magnetized near the film surface. $M_z$ is consisting of basic domains alternately magnetized up and down along the $z$-axis concerning the film surface.

II) $H$//SD and $H$↑//$M$↑ (figure 6a and 7a), $H$ is applied along the +$x$-axis.

i) A small $H$ (0~400 Oe)is applied to SD. With the increasing $H$, the shape of the stripe remains nearly unchanged, but the magnetizations at the edge of the stripe and Bloch-type domain walls are rotating to the +$x$-axis. The magnetizations near the film surface and inside the flux closure cap start rotating to the field direction. When $H$ is 400 Oe, the magnetizations near film surface point to +$x$-axis, and the region of flux closure cap reduces a little.

ii) A moderate $H$ (400~1000 Oe) is applied to SD. It can be seen the SD pattern becomes weak. Most of the marginal magnetizations in basic domains and flux closure cap are moving towards the direction of the magnetic field with the increasing $H$, and further cause an increasing region of in-plane magnetizations along the +$x$-axis near the film surface. The number of magnetizations in the critical area between the stripe and Bloch-type domain walls is turning to +$x$ axis with the increasing $H$. When $H$ reaches 1000 Oe, the magnetizations at the edge of basic domains and flux closure are orientating to the +$x$-axis.

iii) A large $H$ (1000~2500 Oe) is applied to SD. All magnetizations are rotating to the +$x$-axis with the increasing $H$. The SD splits into several narrow-weak SD by the increasing number of Bloch-like domain walls at 1500 Oe, and disappears gradually when the film is in the saturating state (2500 Oe). The magnetization of the film is finally aligning to the +$x$ axis.

III) $H$//SD and $H$↓//$M$↑ (figure 6a and 7a), $H$ is applied along –$x$-axis.

i) A small $H$ (0~400 Oe) is applied. With the increasing $H$, the basic domains of the stripe remain nearly unchanged. The tilted magnetizations near the film surface and inside the flux closure cap as well as the boundary of Bloch-type domain walls are gradually orientating to the field direction. When $H$ is 400 Oe, part of magnetizations in the critical area between the basic domains and Bloch-type domain walls are reversed to –$x$-axis. The regions of flux closure cap and Bloch-type domain walls decrease obviously due to the rotation of the magnetization at the edge of flux closure cap and Bloch-type domain walls to the direction of the magnetic field with the increasing $H$. All the tilted magnetizations near the film surface orientate to $-x$-axis.

ii) A moderate $H$ (400~1000 Oe) is applied to SD. The SD pattern becomes weak, and the marginal magnetizations in basic domains are rotating to the direction of the magnetic field with the increasing $H$. The regions of flux closure cap further reduce, and only a small flux closure core is observed when $H$ reaches 1000 Oe. This further causes an increasing proportion of magnetization paralleling to the $-x$-axis near the film surface. The magnetizations in the critical area between the basic domains and Bloch-type domain walls are firstly orientated to $-x$-axis. Part of the magnetizations along $+x$-axis in Bloch-type domain walls change to be along the $-x$-axis at 500 Oe, and completely turn towards $-x$-axis when $H$ exceeds 600 Oe.

iii) A large $H$ (1000~2500 Oe) is applied to SD. All magnetizations are rotating to $-x$-axis with the increasing $H$. The SD splits into several narrow -weak SD by the increasing number of Bloch-like domain walls at 1500 Oe and disappears gradually when the film is in the saturating state (2500 Oe). The magnetization of the film is finally aligned to $-x$-axis.

IV) $H\perp$SD and $H\uparrow\perp M\uparrow$, $H$ is applied along the $+y$ axis.

i) A small $H$ (0~400 Oe) is applied. The magnetizations in the critical area between the basic domains and Bloch-type domain walls show a little deviation along the direction of the magnetic field, and become more obvious with the increasing $H$. The flux closure caps parallel (antiparallel) to the field direction expand (shrink) slowly with the increasing $H$, this results in the component of the $M_y$ parallel to the field. $M_y$ linearly increases with $H$, which is in agreement with the magnetization curve measurement (black square line in figure 2). The centers of Bloch-type domain walls were found to shift alternately upwards and downwards along the $z$-axis. The above variation of the flux closure caps and Bloch-type domain walls causes the twist of basic domains and magnetization near the film surface.

ii) A moderate $H$ (400~1000 Oe) is applied to SD. The expanding or shrinking of flux closure caps, the shifting of the centers of Bloch-type domain walls, and the twist of basic domains and magnetization near film surface are more obvious at 500 Oe. The SD is rotating towards the applied field direction ($+y$) when the intensity of the $H$ is 600 Oe. The magnetization at this moment is similar to the simulation result of $H$//SD except for the direction of SD. The magnetizations are consisting of a pair of reversed magnetizations at the film surface along the $\pm x$-axis and the tilted magnetizations along the $+y$-axis. When the $H$ is further increased, the SD is completely turning to the $+y$-axis.

iii) A large $H$ (1000~2500 Oe) is applied to SD. The results are similar to the simulation results of $H$//SD. All magnetizations are rotating to the $+y$-axis with the increasing $H$. The SD splits into several narrow-weak SD by the increasing number of Bloch-like domain walls at 1500 Oe, and disappears gradually when the film is in the saturating state at (2500 Oe). The magnetization of the film is finally aligned to the $+y$-axis.

The micromagnetic simulation results can clearly explain the behavior of magnetization curves (figure 2):

i) For $H$//SD and $H\uparrow//M\uparrow$, the component of the magnetization paralleling to the external field increases linearly until its saturation state, which is in agreement with the magnetization curves (red circle line in figure 2).

ii) For $H$//SD and $H\downarrow//M\uparrow$, the magnetization is originally opposite with the direction of the field, and then part of the magnetization is rotating towards the field direction. This rotation of magnetization begins to strongly increase when $H$ exceeds 500 Oe (corresponding to the $H_c$ in figure 2), and further displays the same magnetization state with $H\uparrow//M\uparrow$ when $H$ is larger than 1000 Oe (corresponding to the $H_{rev}$ in figure 2). Finally, the curves are getting to their saturation state.

iii) For $H\perp$SD and $H\uparrow\perp M\uparrow$, the initial magnetization is zero due to the equal and negative magnetization component $M_y$ at remanence. When the $H$ is increasing, the component of the magnetization paralleling to the external field increases until the stripes rotate towards the field direction at 600 Oe (corresponding to the $H_{tra}$ in figure 2). When $H$ further increases, the magnetizations then change similarly with $H$//SD, and reach the saturation state.

We note that the higher value of the reorientation field found in the micromagnetic simulations, concerning the experiment, maybe due to edge effects induced by the limited extension of the simulated cell [24, 46]. We performed a large number of simulations by changing the thickness, saturation magnetization, perpendicular anisotropy, exchange coefficient, etc. (not shown). The results show that these parameters affect the domain width, critical thickness, saturation field, etc. of the SD, but the physical process and mechanism of this work are not affected. The domain width $w$ is reduced with the increased applied field $H$, and previous research has studied this systematically [30, 60], thus we do not discuss it here.

### D. Discussion and calculation

The underlying equation of motion for the temporal evolution of the magnetization is the Landau–Lifschitz–Gilbert (LLG) equation [14, 61]:

$$\frac{d\vec{m}}{dt} = -|\gamma|\vec{m}\times\vec{H}_{eff} + \alpha\vec{m}\times\frac{d\vec{m}}{dt} \qquad (1)$$

where $\boldsymbol{m}$ is the magnetization direction, $\gamma$ is the gyromagnetic ratio, the dimensionless coefficient $\alpha$ is called the Gilbert damping constant, and $\boldsymbol{H}_{eff}$ is the effective magnetic field including the external, demagnetization, and anisotropy fields. To excite the acoustic and optic modes, certain experimental conditions have to be met which can be summarized by the following two general rules [31]:

i) The pumping field $\boldsymbol{h}_{rf}$ must have a nonzero component perpendicular to the static magnetization $\boldsymbol{M}$, to exert a finite torque on the magnetization and tilt it out of its equilibrium position:

$$\vec{M} \times \vec{h}_{rf} \neq 0 \quad (2)$$

ii) The total dynamic moment $\boldsymbol{m}_{tot}$ must have a nonzero projection parallel to the direction of the pumping field $\boldsymbol{h}_{rf}$:

$$\vec{m}_{tot} \cdot \vec{h}_{rf} \neq 0 \quad (3)$$

These magnetic moment processions of each excitation mode can finally attribute to the local or total effective field in the film, and one can describe the modes by their resonance frequency. First, combining the results of micromagnetic simulation (figure 6 and figure 7) and magnetization curves (figure 2), it can be sure that the results of the different resonance frequencies and modes are highly dependent on the change of the magnetization and SD. The detailed conclusion is below:

i) For $h_{rf}\perp$SD, it is the AM. The magnetic moments in the neighbor stripe resonate in-phase and are coupled by the domain surface charges. This gives rise to the dynamic dipolar coupling field [31]. Such dynamic dipolar fields add to an out-of-plane restoring torque and induce the precession frequency. It can be considered that there is a pseudo-anisotropy along the direction of SD. The permeability spectra (figure 3b and 3d, figure 4a-b and 4d-e) are in accordance with the conventional in-plane uniform precession mode. We note that the permeability spectra of figure 3 (b, d) and figure 4 (a, b, d, and e) are substantially the same. We list the results of figure 4 to discuss here. It is worth noting for the situation $h_{rf}\perp$SD and $H\downarrow//M\uparrow$ that the $f_r$ reduces first due to the opposite initial direction between $H$ and $M$. The $M$ rotates to the direction of $H$ gradually with the increasing $H$ (figure 6b and 7b). When $H$ is larger than $H_c$, most $M$ turns towards the direction of $H$. The total $M$ is further orientating along the direction of $H$ as the $H$ exceeds $H_{rev}$. We note that when the magnetic field exceeds $H_{rev}$, the magnetization along the field direction in SD will linearly increase i.e., the in-plane magnetization curves of $H\uparrow//M\uparrow$ and $H\downarrow//M\uparrow$ are overlapping (figure 2). This conversion from $H_c$ to $H_{rev}$ is very quick, corresponding to the short inflection in the permeability spectra (figure 4b and 4e). The $f_r$ of $H\downarrow//M\uparrow$ then increases linearly and shows the same law with that of $H\uparrow//M\uparrow$ after $H$ is larger than $H_{rev}$.

ii) For $h_{rf}$//SD, this is the OM. The magnetic moments in the neighbor stripe resonate out of phase and are coupled by the wall surface charges, which give rise to the in-plane dynamic dipolar coupling field [31]. The dynamic dipolar fields add to an in-plane restoring torque and enhance the resonance frequency. Thus, the $f_r$ of the OM is higher than AM. When a small $H$ is applied perpendicular to SD, although the $M$ is disturbed towards the direction of $H$ (figure 6c and 7c), the SD does not rotate. The $h_{rf}$ is still orthogonal to SD, the film also resonates with the OM (figure 4c and 4f). With the further increasing $H$, SD begins turning to the direction of $H$, the OM weakens while the AM appears. The resonance peaks in the field range of 90 to 150 Oe become complicated and the field at this range ($H_{tra}$) is also the transition value of SD rotation. These multiple peaks attribute to the domain, domain wall, and the flux closure cap, which have been calculated detailedly by Ebels *et al* and Vukadinovic *et al* [4, 14, 31]. When the SD orientates completely to the direction of $H$, the OM and AM will compose to AM, finally continues as the uniform precession mode.

To describe quantitatively the relation of the resonance frequency depending on the $h_{rf}$ and $H$, we further demonstrate the resonance equation of different modes. As well know, the resonance frequency $f_r$ of the film can be determined by Kittel equation [62]:

$$f_r = \frac{\gamma}{2\pi}\sqrt{\left(H_{eff} + 4\pi M_s\right)H_{eff}} \tag{4}$$

However, the SD film is in-plane isotropy, and the presence of the response in SD can be related to its local magnetization, where the magnetizations are alternatively up and down in the stripes. Thus, the contribution of the effective magnetic field $H_{eff}$ of SD is complicated, and its magnetization is not a constant of $4\pi M_s$, but changes as a function of $H$ until saturation. The Kittel formula needs to be revised. According to previous results [33, 42, 43, 63], the free energy density of periodic domains consists of magnetostatic energy, anisotropy energy, exchange energy, and magnetic field energy, and the frequencies of the excitation modes are calculated by the Smit-Beljers procedure [40, 42, 43, 63, 64]. The resonance frequency of AM and OM with applied magnetic field $H$ is rewritten as follow:

i) $h_{rf}\perp$SD, $H$//SD, and $H\uparrow//M\uparrow$

$$f_r^{AM} = \frac{\gamma}{2\pi}[(H + H_k^{dyn} + 4\pi M)(H + H_k^{dyn})]^{1/2} \tag{5a}$$

$$f_r^{AM,SW} = \frac{\gamma}{2\pi}[(H + H_k^{dyn} + 4\pi M + H_{sw})(H + H_k^{dyn} + H_{sw})]^{1/2} \tag{5b}$$

ii) $h_{rf}\perp$SD, $H$//SD and $H\downarrow//M\uparrow$

When $H<H_c$,

$$f_r^{AM} = \frac{\gamma}{2\pi}[(H_k^{dyn} - H + 4\pi M)(H_k^{dyn} - H)]^{1/2} \tag{6a}$$

$$f_r^{AM,SW} = \frac{\gamma}{2\pi}[(H_k^{dyn} - H + 4\pi M + H_{sw})(H_k^{dyn} - H + H_{sw})]^{1/2} \tag{6b}$$

When $H>H_c$, the equations of $f_r^{AM}$ and $f_r^{AM,SW}$ are the same as Eqs. (5a) and (5b) respectively.

iii) $h_{rf}$//SD, $H\perp$SD and $H\uparrow//M\uparrow$ or $H\downarrow//M\uparrow$The equations of $f_r^{AM}$ and $f_r^{AM,SW}$ are the same as Eqs. (5a) and (5b) respectively.

$$f_r^{OM} = \frac{\gamma}{2\pi}[(H + H_k^{dyn} + H_{ex}^{SD} + 4\pi M)(H + H_k^{dyn} + H_{ex}^{SD})]^{1/2} \tag{7a}$$

$$f_r^{OM,SW} = \frac{\gamma}{2\pi}[(H + H_k^{dyn} + H_{ex}^{SD} + 4\pi M + H_{sw})(H + H_k^{dyn} + H_{ex}^{SD} + H_{sw})]^{1/2} \tag{7b}$$

Where,

$$H_{sw} = \frac{2A}{M}\left(\frac{n\pi}{D}\right)^2 \tag{8}$$

$H_k^{dyn}$: the dynamic anisotropy of SD. This represents the sum of the inner anisotropy of SD film including rotational anisotropy of SD and static anisotropy of the film. In this work, we did not induce a static anisotropy during the deposition, and this term can be neglected.

$H_{sw}$: the spin-wave field of SD. The spin wave comes from the standing wave resonance between periodic SD.

$n$: the quantization number SW.

$D$: the period of SD.

$H_{ex}^{SD}$: the exchange coupling field between the adjacent stripes.

$M$: the magnetization of film. It can be obtained from the magnetic hysteresis loops or magnetization curves. The values of $M$ are depending on the $H$, and the results are shown in figure 8(a-b). $M_s$ is the saturation magnetization.

Based on the above Eqs. (5-8), we obtained the resonance frequency $f_r$ at each magnetic field $H$ by VNA-FMR, the corresponding $4\pi M$ at each magnetic field obtained from VSM and magnetization curve. Finally, the $H_k$ dependent magnetic field can be estimated. The calculated $H_k^{dyn}$ and $H_k$ are presented in figure 8 (c-f). The physical process of the measured two modes or measured three situations have been discussed above, and the change of the corresponding different $H_k$ fits well with these discussions. We can obtain that the variation of the $f_r$ of different resonant modes with external magnetic field $H$ is not only highly related to magnetization distribution but also the SD induced anisotropy (i.e., the direction of SD). The values of SD-induced anisotropy are dependent on the strength and direction of SD, which changes with the $H$. In addition, the transition/rotation field ($H_{tra}$) of the SD is more hysteresis than the reversal field ($H_{rev}$) of the magnetization. It can be seen that the value of $H_{tra}$ is about 60 Oe larger than $H_c$ and 30 Oe larger than $H_{rev}$ for both sputtered and electrodeposited films. In addition, the stronger exchange coupling anisotropy in the sputtered SD film than the electrodeposited SD film proves the more regular stripe in sputtered film, which agrees with the above MFM results.

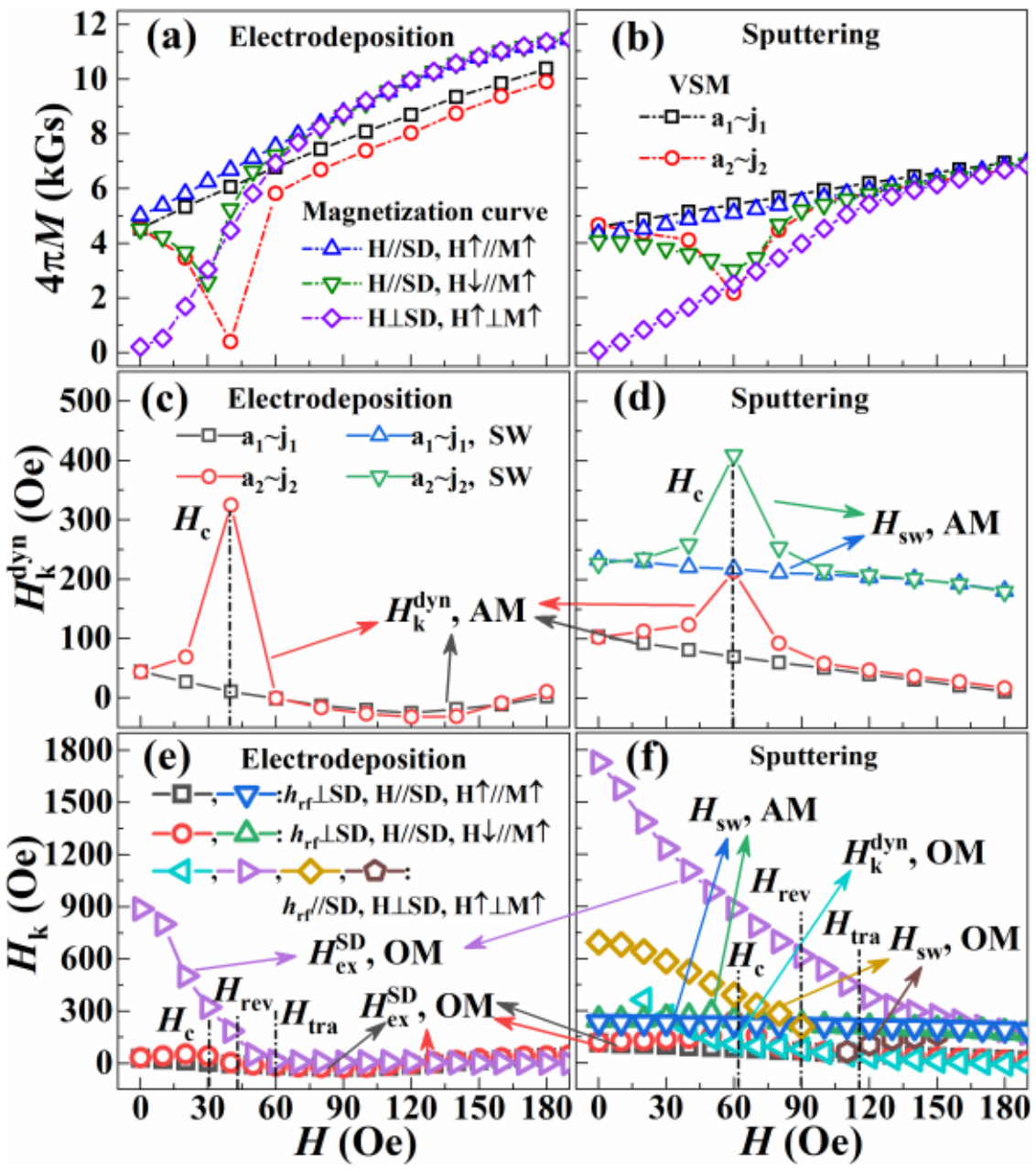

Figure 8. The magnetizations $4\pi M$ (a-b) and different anisotropy $H_k$ (c-f) of different resonant modes depend on the applied magnetic field $H$. In (a-b), the $4\pi M$ are the experimental result obtained by VSM and magnetization curves. In (c-f), the result is obtained by using Eqs. (5-8).

As can be seen in the magnetic spectra of figure 3 and 4, the peak intensity of imaginary permeability are not equal for the same film with the different applied magnetic fields. We further calculate the results and compare them with the experimental results.

Archer *et al* [65] has demonstrated the limitation between $f_r$ and the initial

permeability $\mu_{in}$:

$$(\mu_{in}-1)f_r^2=(\frac{\gamma}{2\pi})^2(4\pi M_s)^2 \tag{9}$$

The maximum intensity of imaginary permeability peak $\mu''_{max}$ can be obtained by solved LLG equation [66]:

$$\mu''_{\max}=\frac{1}{2}(\mu_{in}-1)\sqrt{1+\frac{1}{\alpha^2}} \tag{10}$$

$\mu''_{max}$ finally turns out for SD to be the form

$$\mu''_{\max}=\frac{(4\pi M)^2}{2[(H+H_k^{dyn})(4\pi M+H+H_k^{dyn})]}\sqrt{1+\frac{1}{\alpha^2}} \tag{11}$$

It can be seen from Eq. (11), the $\mu''_{max}$ depends on $\alpha$, $4\pi M$, and $H_k^{dyn}$. For the same sample, the $\alpha$ is a constant, which is fitting as 0.01 for sputtering film and 0.07 for electrodeposition film in this work. The calculative results of $\mu''_{max}$ are shown in figure 9. It can be seen that changes in the measurement and calculation data are in good agreement. This means that the modified equations are suitable to explain the change process of SD resonance intensity and reflect the result of $\mu''_{max}$ under the increased $H$. The results indicate that the resonance intensity of different modes in SD is determined by the local magnetization rather than the whole magnetization of SD. We note that the result of the electrodeposition film and the result of sputtering film in the field range of 90 to 150 Oe do not fit very well. This is due to the large error of OM for electrodeposition film (the resonance intensity is low, the line width is very large, and we have explained the reason above) and the low distinguishing of the overlapping peaks n the field range of 90 to 150 Oe.

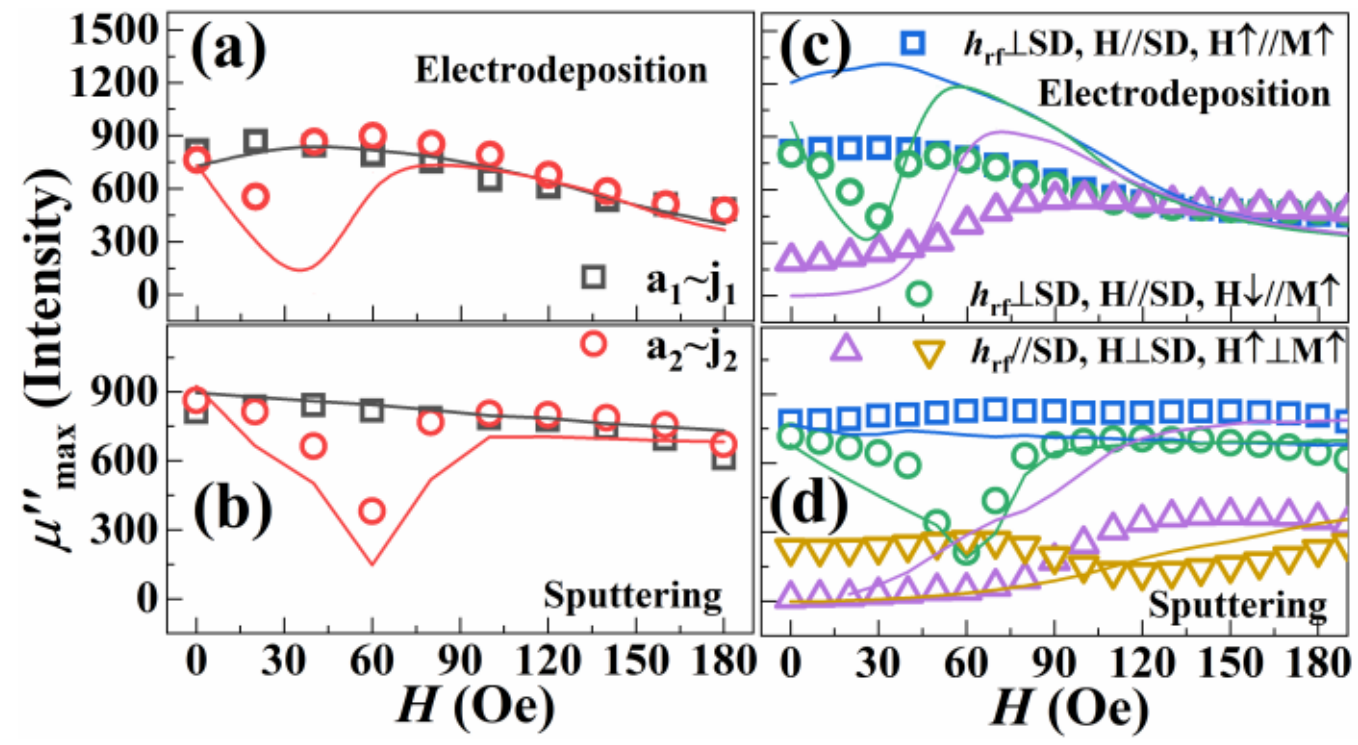

Figure 9. Experimental and calculative maximum intensity of imaginary permeability peak $\mu''_{max}$. The different shape is the experimental result while the line is the calculative result. The experimental result and calculative result are in the same color. Figure (a-b) corresponds to the resonance intensity of figure 3, and Figure (c-d) corresponds to the resonance intensity of figure 4.

## 4. Conclusion

In summary, we first investigated the structure and magnetic domain of FeNi SD films. It is found that the electrodeposited film showed a dispersive SD pattern and resulted in a weak exchange coupling interaction when compared with sputtered film. The static magnetic properties especially the magnetization curve reveal the magnetization distribution in SD was dependent on the direction of SD. The magnetization starts to reversal after the applied field exceeds $H_c$. The rotation of SD is hysteresis with 60 Oe larger magnetic field than $H_c$. Such magnetization and SD distribution determined the selectively excited dynamic microwave magnetic properties, which emerged the dynamic hysteresis, the AM, OM, and SW resonance.

The frequency and intensity of different resonance modes of stripe domain are determined by the local magnetization. The SD rotation and magnetization reversal were further certified by the micromagnetic simulation. Based on the above results, the anisotropy and resonance intensity of different modes were calculated by the modified resonance equations, and the results fit well with the experimental data. The results help to deeply understand the rotation mechanism of the SD and provide the possibility of SD film for microwave excitation applications in spintronics.

**Acknowledgments**

This work is supported by the National Natural Science Foundation of China (11704211 and 11847233), China Postdoctoral Science Foundation (2018M632608), the Applied basic research project of Qingdao (18-2-2-16-jcb), the basic scientific research business expenses of the central university, and Open Project of Key Laboratory for Magnetism and Magnetic Materials of the Ministry of Education, Lanzhou University.